\documentclass[twocolumn,aps,pra,a4paper,showpacs,preprintnumbers,amsmath,amssymb]{revtex4}

\usepackage[english]{babel}
\usepackage[dvips]{graphicx}
\usepackage{dcolumn}
\usepackage{bm}
\usepackage{amssymb}
\usepackage{ulem}
\usepackage{mathrsfs}
\begin{document}
\title{Reconstructing vibrational states in warm molecules using four-wave mixing with femtosecond laser pulses}

\author{Anders S. Mouritzen}
\email{asm@phys.au.dk} \affiliation{Lundbeck Foundation Theoretical
Center for Quantum System Research, Department of Physics and
Astronomy, University of Aarhus, DK-8000 \AA rhus C, Denmark}

\author{Olga Smirnova}
\email{olga@ned.sims.nrc.ca} \affiliation{NRC Canada, 100 Sussex
Drive, Ottawa, Ontario K1A 0R6 Canada}

\author{Stefanie Gr\" afe}
\email{Steffi.Graefe@.nrc.ca} \affiliation{NRC Canada, 100 Sussex
Drive, Ottawa, Ontario K1A 0R6 Canada}

\author{Rune Lausten}
\affiliation{NRC Canada, 100 Sussex Drive, Ottawa, Ontario K1A 0R6
Canada}

\author{Albert Stolow}
\affiliation{NRC Canada, 100 Sussex Drive, Ottawa, Ontario K1A 0R6
Canada}

\date{\today}

\begin{abstract}
We propose a method to reconstruct the vibrational quantum state of
molecules excited by a general excitation laser pulse. Unlike
existing methods, we do not require the molecules before excitation
to be in a pure state, allowing us to treat the important case of
initially thermally excited molecules. Even if only a single initial
level is appreciably populated, initial levels with small
populations can still give major contributions to the unknown
vibrational state, making it essential to take them into account. In
addition to the excitation pulse, the method uses two incident,
short laser pulses in a non-co-linear geometry to create four-wave
mixing in the molecules. The measurements used in the reconstruction
are spectra of the outgoing four-wave mixing pulse at different time
delays of the excitation laser pulse. An important point is that the
method does not require detailed knowledge of molecular transition
moments between excited states nor of any of the incoming laser
pulses, but circumvents this requirement by using one or more
calibration laser pulses in a separate experiment either before or
after the main data are recorded. The only requirements for the
calibration laser pulses are that the constant parts of their
spectrums should together cover the spectral range of the excitation
laser pulse, and the constant part of each should have sufficient
spectral overlap with one other calibration pulse to populate two of
the same levels. Finally, we discuss the extension of the
reconstruction method in this paper to more general situations,
hereby presenting the new idea of quantum state reconstruction
through perturbations with calibration.
\end{abstract}



\maketitle

\section{\label{sec:intro}Introduction}
Everything worth knowing about a molecule at any given time is its
physical state. Hence, whether one's interest lies in interaction of
radiation with the molecule, chemical reactivity or intra-molecular
processes, the state contains this information. Evidently, it is of
great interest to be able to determine the state of a molecule. In
this theoretical paper, we show how one can determine a molecular
state in a four-wave mixing experiment. The method presented is
particularly useful for determining the molecular states prepared by
a single, optically tailored femtosecond laser pulse. Production of
this type of laser pulses has been demonstrated
\cite{tailorWeiner1}-\cite{tailorBucksbaum} and they have been used
to produce specific vibrational states in molecules
\cite{tailorYan}, \cite{tailorAssion}.

Finding the state within a quantum mechanical description is
complicated, as there is no single observable giving the quantum
state. This is well-known for pure states, i.e. states that can be
characterized by a single wave function, where there is no
observable directly revealing this complex wave function. The same
absence is true for the more general case of mixed states, which are
states that can be characterized by a density matrix, but not
necessarily by a single wave function. In this paper, we shall be
treating such mixed states, paying special attention to thermal
states; a type of mixed state that describes a quantum system at a
finite temperature. In particular, we will concentrate on finding
the vibrational state of an initially thermal molecule, which has
been excited to an unknown vibrational state by a short laser pulse.

Although the state is not directly observable, one can determine the
state from measurements. This is the field of study known as quantum
state reconstruction. On each member of an \textit{ensemble}, one or
more commuting measurements are performed, whereafter this member is
discarded. This ensemble must be large enough that the measured
values are close to their true expectation values, which obviously
requires numerous measurements of each quantity. Fortunately, in the
method presented below there will be a vast number of molecules in
each experimental run, ensuring excellent statistics.


In the broader picture, there has recently been growing interest in
reconstructing quantum states for various systems, for a review see
\cite{welschlangtreview}. The general problem of quantum state
reconstruction is experimentally and theoretically challenging, and
methods tend to apply to only a very particular setting. The most
well-known is the tomographic reconstruction of harmonic oscillator
states from measurements of spatial distributions \cite{invradon},
\cite{invradon2}. Other systems include particles in traps (neutral
atoms \cite{Buzekatom} and ions \cite{Wineland}), general
one-dimensional systems \cite{LeonhardtRaymerprl}, dissociating
diatomic molecules \cite{Juhl} and the angular state of a
dynamically aligned molecule \cite{rotafos}.

Earlier works directly related to this paper have proposed
reconstructing vibrational states for diatomic molecules using
heterodyne detection of fluorescence \cite{Heterodyndiatom}, for
general molecules using time- and frequency- resolved fluorescence
\cite{Shapiro1995}, \cite{Shapiro1996}, and using time- and
frequency-integrated fluorescence and a known reference state for
quantum state holography \cite{albholografi}. Common for these
proposals is the requirement that the initial state be a pure state.
Realistically, fulfilling this requirement in turn demands a
sufficiently low temperature of the molecules before the creation of
the vibrational excitation. This initial temperature must be so low
that only the lowest vibrational level is populated. For many
molecules this is unpractical, e.g. gaseous iodine, where practical
temperatures populates more than one vibrational state, lest the
iodine is deposited \cite{tailorYan}, \cite{Runeogalbert}. It is
important to note that even though the initial thermal state's
ground vibrational state population is usually much larger than that
of all other vibrational states, it does not guarantee that this
gives the dominating contribution to the unknown excited state. For
instance, in the experiment on molecular iodine described in
\cite{Runeogalbert}, the ground and first excited vibrational level
in the initial thermal state give approximately equal contributions
to the unknown excited state, even though the ratio of populations
is about $2.5:1$. This is mainly due to the different Franck-Condon
factors involved in the formation of the unknown excited state by
the first laser pulse. Hence, taking several initially thermally
populated levels into account can be essential to understanding the
excited state formed.

The paper is arranged as follows: In section~\ref{sec:stme} we
identify the type of quantum state being reconstructed and give a
brief description of what is required of the measurements. In
section~\ref{sec:experim} we outline an experimental implementation
of the required measurements and give an overview of the energy- and
time-regimes involved. In section~\ref{sec:teori} we show by
calculation how to reconstruct the unknown vibrational quantum
state. In section~\ref{sec:disk} we discuss the generality of the
reconstruction procedure and conclude the paper.

\section{States and measurements\label{sec:stme}}
Having established the importance of the physical state and its
reconstruction, we will now specify what we mean by the physical
state and what the measurements for reconstruction must fulfill.

It is essential to distinguish between the state of a system and its
dynamics. The state contains the answer to all questions we may ask
the system at a certain point in time, i.e. the result of all
thinkable one-time measurements at this time. In sharp contrast to
this state is the concept of dynamics, which is prescriptions for
finding the state at a later time, from knowing it at an earlier
time.

In the subject of state reconstruction, and in this paper in
particular, the goal is to determine the state of a system. The type
of state we will find in this paper is an unknown vibrational state
of a sample of molecules, coherently excited from an initial thermal
state. Specifically, we will restrict the reconstruction of the
unknown vibrational state to the electronic state(s) it populates,
denoted by the electronic index $a$, and thus not find correlations
with the initial thermal population. This is exactly what is usually
meant by the concept of an excited vibrational state. For
transparency, we restrict ourselves to the most important case where
the initial thermal state is limited to the electronic ground state,
with possibly several vibrational levels populated. The excitation
that forms the unknown vibrational state need not be perturbative in
the sense that it may transfer a large fraction of the initial
thermally populated level to the electronic state $a$. Nevertheless,
it must leave a non-vanishing population in the initial state and
not transfer population back to other of the initially populated
levels.

Since the initial state is a mixed state, we cannot describe it by a
single wave function, and because coherent excitations preserve the
mixed character, neither can we describe the excited vibrational
state this way \cite{Grimberg}. A sufficing description is the
density matrix, for which the initial thermal state would be
diagonal in the energy basis, but we shall take a more transparent
course.

As the states originating in different energy levels in the initial
thermal state are incoherent with respect to each other, and the
vibrational state we seek is coherently excited from these, we can
describe the system as a incoherent sum of vector states
(corresponding to wave functions). By an incoherent sum, we mean
forming a sum of the states, each with a phase factor
$\exp(i\theta_k)$ labeled after the original energy level $k$ in the
initial thermal state from which the coherent excitation took place.
These phase factors will be averaged over when taking inner
products, making incoherent cross-terms vanish. In this paper, each
incoherent term will be labeled by a left subscript, such as
${}_{k}\beta$. To illustrate this, consider the example of a thermal
state of a molecule with the only energy levels populated being the
vibrational levels $k = 0$ and $k = 1$ in the electronic ground
state. We then imagine exciting a vibrational state in another
electronic level $a$, the vibrational eigenstates
$|\phi_{a,l}\rangle$ herein being labeled by index $l$. The excited
state will then be:
\begin{eqnarray}
|\psi_a\rangle &=& \sum_k \left|{}_k\psi_a \right\rangle \nonumber \\
&=& \left|{}_{k=0}\psi_a \right\rangle e^{i\theta_0} +
\left|{}_{k=1}\psi_a \right\rangle e^{i\theta_1} \nonumber \\
&=& e^{i\theta_0}\sum_l {}_{0}\beta_{l}\left|\phi_{a,l}
\right\rangle + e^{i\theta_1} \sum_l {}_{1}\beta_{l}\left|\phi_{a,l}
\right\rangle, \nonumber
\end{eqnarray}
where the ${}_{k}\beta_l$ are expansion coefficients. What we mean
by ``finding the unknown vibrational state" is thus determining all
the complex expansion coefficients ${}_{k}\beta_l$. With the
incoherent averaging, inner products between states originating from
different levels in the initial thermal state vanish, e.g.
\begin{eqnarray}
\langle {}_{k=0}\psi_a |\hat{A}| {}_{k=1}\psi_a \rangle &\propto&
\int_0^{2\pi}\hspace{-0.35cm}d\theta_0\, e^{-i\theta_0}
\int_0^{2\pi} \hspace{-0.35cm} d\theta_1\, e^{i\theta_1} =
0,\nonumber
\end{eqnarray}
where $\hat A$ can be any operator. Thus, we will not see any
cross-terms between states originating in different energy levels of
the initial thermal state, whence we will conveniently treat these
states for different $k$ separately. However, one should notice that
it is still possible to see interference in an intensity signal
between the light formed from different incoherent levels. This can
be easily seen by considering the two electric fields formed from
the transitions from level $a$ to the electronic ground state $0$:
\begin{eqnarray}
{}_{k=0}E(t) &\propto& d^2/dt^2 \langle {}_{k=0}\psi_0 |\hat{d}|
{}_{k=0}\psi_a \rangle\nonumber\\
{}_{k=1}E(t) &\propto& d^2/dt^2 \langle {}_{k=1}\psi_0 |\hat{d}|
{}_{k=1}\psi_a \rangle,\nonumber
\end{eqnarray}
where $\hat{d}$ is the scalar product of the dipole moment operator
with the electric field's polarization vector. Since the incoherent
$\theta_k$-factors cancel in each electric field, we will observe
interference in the intensity signal $I \propto \left[{}_{k=0}E(t) +
{}_{k=1}E(t)\right]^2$.

Having accounted for what we mean by the state, we now turn our
attention to measurements. To accomplish the state reconstruction,
we will require knowledge of the results of enough measurements to
uniquely determine the state \cite{Scully}. Such a set of
measurements is known as a \textit{quorum}.

Finding a quorum can be straightforward or tedious, depending on
both the system and the level of description. In a classical system
of particles, one quorum is given by measurements of the coordinates
and momenta of all particles at a fixed point of time. In contrast,
the quantum mechanical description used in this paper makes finding
a practically realizable quorum much more challenging.
We will use the prevalent approach which is to perform the same type
of measurement at different points of time, thereby letting the
dynamics reveal the state, see e.g. \cite{invradon},
\cite{LeonhardtRaymerprl}. Using this approach, one is almost always
forced to assume full knowledge of the dynamical laws of the
physical system. However, we shall largely circumvent such
assumptions by instead performing a calibration of the measurement
apparatus. Specifically, we will show that spectra of a four-wave
mixing laser pulse, created at different times after the formation
of an excited vibrational state, encompass a quorum for determining
the set $\left\{{}_{k}\beta_l\right\}$.

\section{Experimental implementation outline \label{sec:experim}}
\begin{figure}[htbp]
\includegraphics[width=0.48\textwidth]{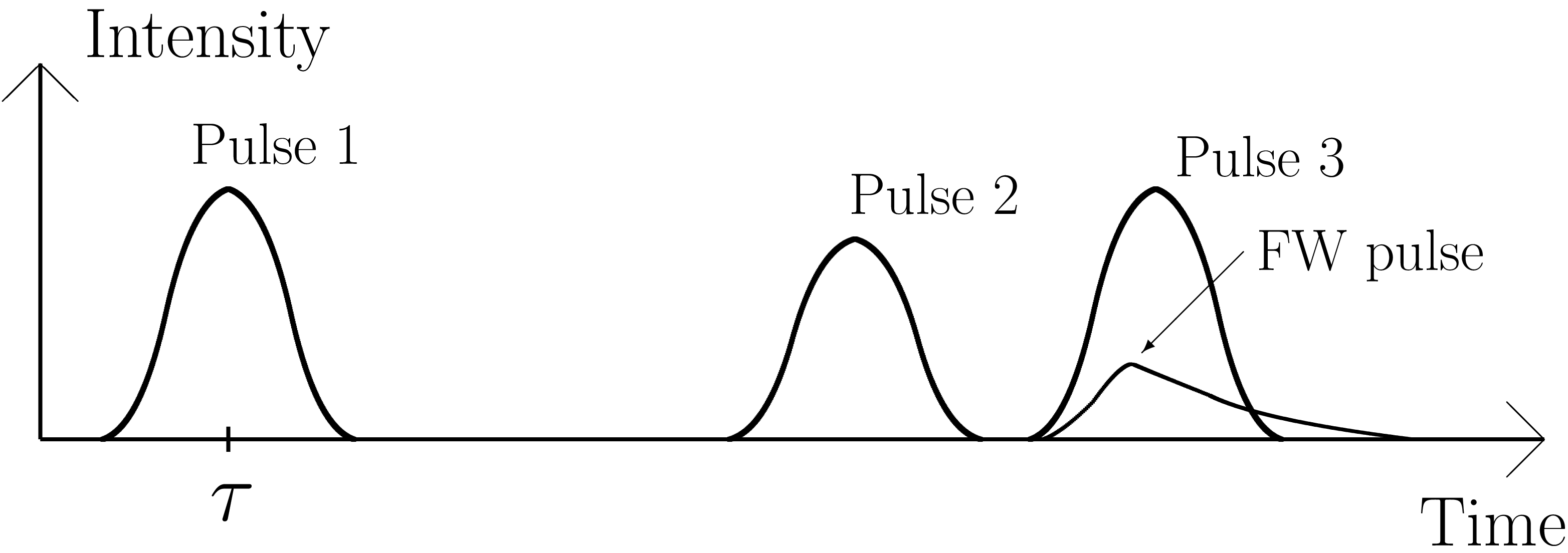}
\caption{Time sequence of the laser pulses. First, pulse~$1$ excites
a vibrational state in a sample of thermal molecules at time $\tau$.
When pulse~$1$ is over, the molecules are probed using the
pulses~$2$ and $3$. While the pulses~$1$-$3$ are always the same,
the time $\tau$ is varied from one experimental run to the next. The
interaction of the three pulses~$1$-$3$ with the molecules gives
rise to a four-wave mixing (FW) pulse. The spectrum of this FW pulse
is measured together with the current value of $\tau$. Of further
notice is that the emission of the coherent FW pulse can continue
long after the incident pulses~$1-3$ have died out, since the
emission is due to an excitation in the molecules. For transparency,
we will in section~\ref{sec:teori} assume that pulse~$2$ and $3$
have no temporal overlap, as shown. However, this is not a
requirement of our method, and we will relax this assumption in
section~\ref{sec:disk}, allowing for overlapping probe pulses as
used in \cite{Runeogalbert}. \label{fig:Toversigt}}
\end{figure}

In this section we outline a pump-probe experiment where one can
record the data necessary for the vibrational state reconstruction.
This is a four-wave mixing experiment where one sends in three laser
pulses~$1$-$3$ from different directions on a molecular sample, and
records the spectrum of the outgoing four-wave mixing (FW) pulse
\cite{NL bible}, \cite{Faeder}. Laser pulse~$1$ reaches the
molecular sample first and can be regarded as a pump pulse, since it
has the effect of creating the unknown vibrational quantum state in
the molecules. Pulses~$2$ and $3$ arrive at the molecules long after
pulse~$1$ is over, and can thus be regarded as probe pulses . The
time $\tau$ when pulse~$1$ creates the unknown state is varied from
one experimental run to the next, retaining the timing of pulses~$2$
and $3$. One then measures the spectrum of the resulting FW pulse.
The time sequence of the four laser pulses is outlined in
Fig.~\ref{fig:Toversigt}. One should notice that even though the
pulses~$1-3$ should individually be the same from one run to the
next, there is no requirement of relative phase stabilization.

To keep the theoretical treatment transparent, we will now introduce
several simplifying assumptions. We emphasize that these are not
formal requirements, and are only included for clarity. Indeed, we
will do away with them in section~\ref{sec:disk}.

\begin{figure*}[htbp]
\includegraphics[width=\textwidth]{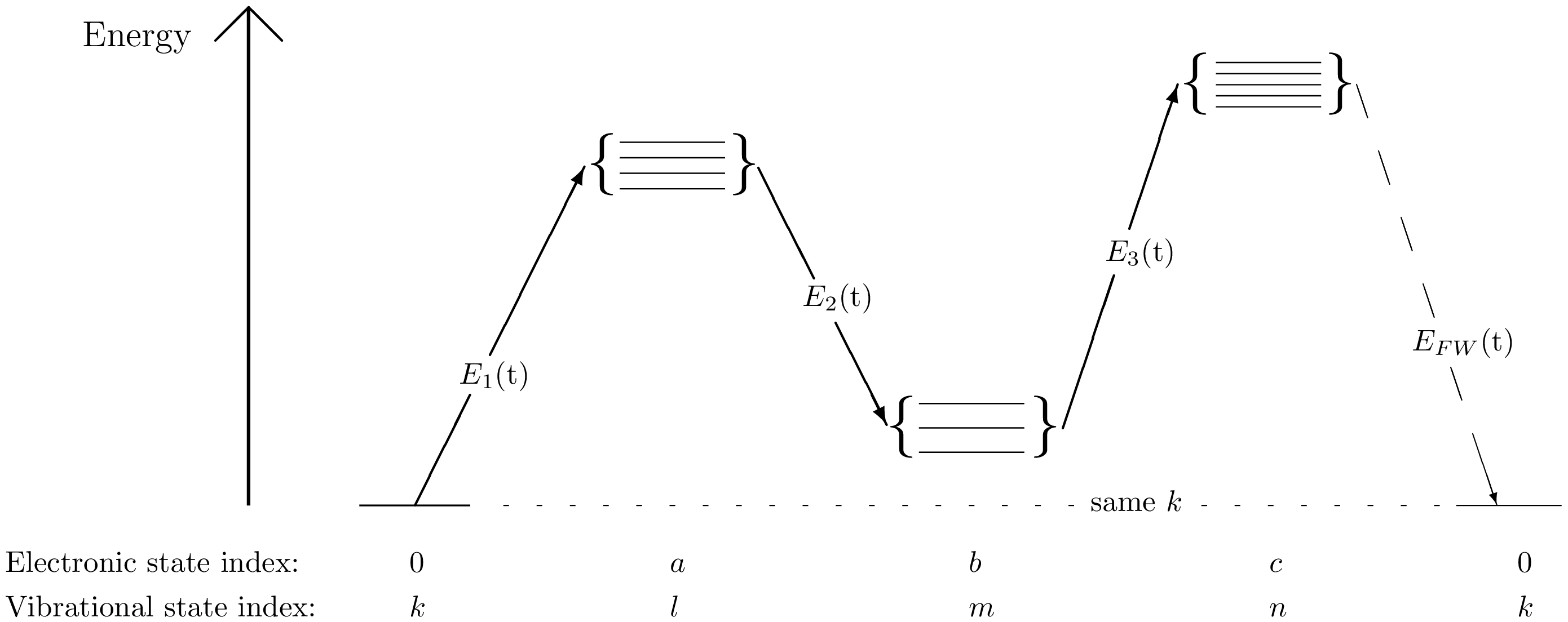}
\caption{Schematic energy diagram showing the levels of the
molecular system and the four laser pulses with time-dependent
electric fields $E_j(t),\, j = {1,2,3,FW}$. Assuming the molecules
are initially thermally excited, we can consider each initially
populated level $k$ in turn, and perform a incoherent sum in the
end. The system starts out in the electronic state $0$, wherefrom
laser pulse~$1$ creates a vibrational state in the electronic state
$a$. After a time $\tau$, which is varied from one experiment to the
next, the state created with pulse~$1$ is probed through a four-wave
mixing process using pulses~$2$ and $3$ to transfer the system
through electronic states $b$ and $c$. Under coherent emission of a
$FW$ photon, the system finally ends up in the $k$ where it
originated. Some of these electronic states may be identical, e.g.
$0 = b$ and $a = c$ in the experiment described in
\cite{Runeogalbert}. The delay time $\tau$ is chosen small enough
for pulse~$1$ to be over when pulse~$2$ and $3$ are applied.
 \label{fig:Eoversigt}}
\end{figure*}

The laser pulses involved will all cause transitions in the
molecules, dominated by dipole transitions. In
Fig.~\ref{fig:Eoversigt} we show an overview of such a series of
transitions, each being accompanied by a change of the electronic
state. We have sketched the four electronic states $0$ and $a$-$c$,
and within each of these several vibrational states.
Furthermore, the figure shows that the transitions due to pulse~$2$
always occurs before transitions due to pulse~$3$. Realistically,
this could be the case if pulse~$2$ precedes pulse~$3$ so that they
have no temporal overlap, as shown in Fig.~\ref{fig:Toversigt}. We
also require that only a single photon from pulses~$2$ and $3$ are
involved in each transition. Finally, we neglect the rotational
degrees of freedom. This constitutes the most transparent situation,
and we shall initially calculate the spectrum of the $FW$ pulse in
section~\ref{sec:teori} using these assumptions. In
section~\ref{sec:disk} we will then discuss how to relax these
non-essential assumptions.

\begin{figure}[htbp]
\includegraphics[width=0.48\textwidth]{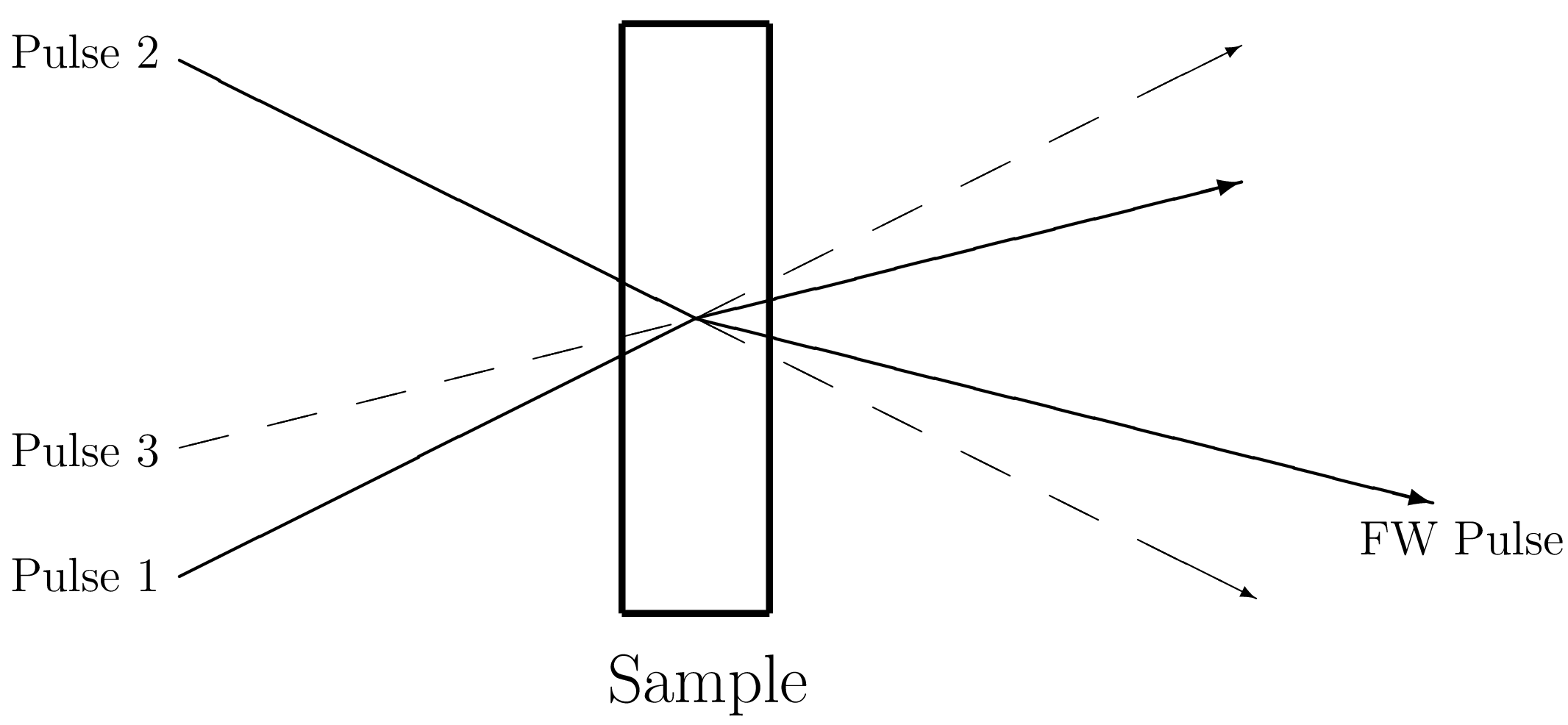}
\caption{Geometric outline of the laser pulses' interactions with
the molecular sample. Dashed lines signify propagation below the
plane of the paper, while fully drawn lines signify propagation
above. It is central to the calculations in this paper that one uses
the shown non-co-linear configuration, in which the pulses~$1$-$3$
have different propagation vectors (BOXCARS-configuration for
pulses~$1$ and $3$ identical). As seen by the molecules, this will
be similar to the pulses~$1$-$3$ approaching from three of the four
corners in a rectangle. By using a pinhole, one may separate out the
FW pulse, hereby ensuring that the measured interaction in the
molecular sample has involved precisely one photon from each of the
pulses~$1$-$3$.
It should be noticed that, in contrast to usual four-wave mixing
situations, the BOX condition gives only state-selectivity of the
electronic state, but not of the vibrational states herein. This
practical inability to tell apart vibrational levels by the angle of
the FW photon arises due to the energetic closeness of the
vibrational levels compared to the energetic distance between the
different electronic states. Thus, transition to a certain
electronic state gives rise to a easily detectable angular
difference, whereas the transition to different vibrational states
herein does not; see also Fig.~\ref{fig:Eoversigt}.
\label{fig:BOXoversigt}}
\end{figure}

One can experimentally justify the implicit assumption made in
Fig.~\ref{fig:Eoversigt} that only a single photon from each of the
incoming laser pulses is involved in each transition. Using the
non-co-linear, so-called folded BOX-configuration shown in
Fig.~\ref{fig:BOXoversigt}, one sends in pulses~$1$-$3$ from three
different directions. Each angle between the pulse propagation
direction is small, typically a few degrees. One can select out the
$FW$ signal caused by precisely one-photon interactions from each
pulse, simply by using pinholes to select out the appropriate
phase-matched
direction for the $FW$ pulse. 

An important point, which will play a key role in the calculations
below, is that the directional selection will typically only select
the number of photons involved in each transition, and thereby the
electronic states, but not select which vibrational levels are
involved in each transition. The reason is that the electronic
states are separated by a comparatively large energy, giving rise to
a realistically observable difference in angle for the $FW$ pulse.
In contrast, the vibrational levels within a certain electronic
level are energetically so close that the focusing angles of the
incoming pulses are much greater than the change in angle needed to
satisfy phase matching to one vibrational state or the other. The
ensuing interference in the $FW$ intensity is precisely the
phase-sensitive quantities that we shall use in the reconstruction.

A final important reason to use the non-co-linear geometry is that
it allows one to observe the $FW$ pulse without the presence of the
very powerful background from pulses~$1$-$3$.

\section{\label{sec:teori}Theory}
In this section, we shall theoretically treat the situation outlined
in the previous section, aiming at reconstructing the quantum state
formed by pulse~$1$. We will accomplish this through calculating the
spectrum of the $FW$ pulse.

We recall from section~\ref{sec:stme} that we initially have a
thermal state, where each energy level $k$ may be treated
independently on the quantum level. Concentrating on a fixed $k$,
the total state formed by the three laser pulses can be written as:
\begin{eqnarray}\label{eq:psitotal}
\Big|{}_k\psi(\tau) \Big\rangle_t &=&
\left|{}_k\psi_0^{(0)}\right\rangle_t +
\left|{}_k\psi_{a_{}}^{(1)}(\tau) \right\rangle_t
+ \nonumber \\
&& \left|{}_k\psi_{b_{}}^{(2)}(\tau) \right\rangle_t +
\left|{}_k\psi_{c_{}}^{(3)}(\tau) \right\rangle_t,
\end{eqnarray}
where the superscript denotes which pulse has formed the state from
the previous, the right subscript denotes the electronic state, and
$t$ is the time. In each of the perturbation orders, we shall for
clarity retain only terms leading to coherent $FW$ emission. The
parameter $\tau$ signifies the time shown in
Fig.~\ref{fig:Toversigt}; the time when the unknown excited state is
formed. This state $\left|{}_k\psi_{a_{}}^{(1)}(\tau) \right\rangle$
does not need to be formed perturbatively, as long as it leaves a
non-vanishing population in the $\left|{}_k\psi_0^{(0)}
\right\rangle$-state, and it does not transfer population back to
other of the thermally populated levels (i.e. other $k$-indices). We
can then expand the unknown state $\left|{}_k\psi_{a_{}}^{(1)}(\tau)
\right\rangle$ on the vibrational eigenstates $|\phi_{a_{},l}
\rangle$ with energy $\hbar \tilde{\omega}_l^a$, belonging to the
electronic level $a$:
\begin{eqnarray}\label{eq:ukendttilst}
\left|{}_{k}\psi_{a_{}}^{(1)}(\tau) \right\rangle_{t_2} &=&
e^{i\theta_{k_{}}}\sum_l {}_{k}\beta_l \left|\phi_{a_{},l}
\right\rangle e^{i \tilde{\omega}_{l,k}^{a_{},0}\tau} e^{-i
\tilde{\omega}_l^a t_2},
\end{eqnarray}
where the ${}_{k}\beta_l$ are expansion coefficients and the
transition frequencies $\tilde{\omega}_{\nu,\nu'}^{\alpha,\alpha'}$
between the vibrational states $\nu$ and $\nu'$ in their respective
electronic states $\alpha$ and $\alpha'$ are given by:
\begin{eqnarray}
\tilde{\omega}_{\nu,\nu'}^{\alpha,\alpha'} &=&
\tilde{\omega}_{\nu}^{\alpha} -
\tilde{\omega}_{\nu'}^{\alpha'}.\nonumber
\end{eqnarray}
Unlike pulse~$1$, the probing pulses~$2$ and $3$ must be
perturbative. Using the electric field $E_2(t)$ of pulse~$2$, we
find:
\begin{eqnarray} \label{eq:psitopert}
\left|{}_{k}\psi^{(2)}_{b}(\tau) \right\rangle_{t_{3}} &=& -\frac{i
q}{\hbar} \int_{-\infty}^{t_{3}} \hspace{-0.47cm} dt_{2}\,
e^{-i \hat H_{b}(t_{3}-t_{2})/\hbar} E_{2}(t_{2}) \nonumber\\
&&\qquad \qquad \times \, \hat d\,
\left|{}_{k}\psi^{(1)}_{a}(\tau) \right\rangle_{t_{2}}, \nonumber\\
\end{eqnarray}
where $\hat H_{\alpha_{}}$ is the vibrational Hamiltonian in the
electronic state $\alpha_{}$, $q = -e$ is the electron's charge, and
$\hat d$ is the dipole moment operator. The corresponding expression
for $\left|{}_{k}\psi^{(3)}_{c}(\tau) \right\rangle_{t}$ can be
found similarly:
\begin{eqnarray}\label{eq:psitrepert}
\left|{}_{k}\psi_{c_{}}^{(3)}(\tau) \right\rangle_t &=&
-\frac{q^2}{\hbar^2} \int_{-\infty}^t \hspace{-0.32cm}dt_3\,
\int_{-\infty}^{t_3} \hspace{-0.32cm} dt_2\,
e^{-i \hat H_{c_{}}(t-t_3)/\hbar} E_{3}(t_3)\nonumber\\
&&\qquad \qquad\times \, \hat{d}\,e^{-i \hat
H_{b_{}}(t_3-t_2)/\hbar}
\,E_{2}(t_2)\,\nonumber\\
&&\qquad \qquad\times \,
\hat{d}\,\left|{}_{k}\psi_{a_{}}^{(1)}(\tau)\right\rangle_{t_2}.
\end{eqnarray}
We introduce the Fourier transforms of the electric fields:
\begin{eqnarray}\label{eq:Efeltfou}
E_r(t_r) &=& \int_{-\infty}^\infty \hspace{-0.32cm} d\omega_r \,
\mathcal{E}_r(\omega_r) e^{-i \omega_r t_r}, \quad r = 2,3.
\end{eqnarray}
Furthermore, we use identities resolved on the vibrational states
$\nu$ of the electronic state $\alpha$, including both bound and
continuum states:
\begin{eqnarray}\label{eq:Identalfa}
\hat I_{\alpha} = \sum_\nu \left|\phi_{\alpha,\nu} \right\rangle
\left\langle \phi_{\alpha,\nu} \right|.
\end{eqnarray}
Finally, we abbreviate $q/\hbar$ times the dipole transition moments
between the vibrational levels $\nu$ and $\nu'$ in their respective
electronic states $\alpha$ and $\alpha'$:
\begin{eqnarray}\label{eq:dipoltrans}
\mathfrak{D}_{\nu,\nu'}^{\alpha,\alpha'} &=& \frac{q}{\hbar}
\left\langle \phi_{\alpha,\nu} \right| \hat{d} \left|
\phi_{\alpha',\nu'} \right\rangle.
\end{eqnarray}
We use Eqs.~(\ref{eq:Efeltfou})-(\ref{eq:dipoltrans}) with
Eq.~(\ref{eq:psitrepert}) and perform the time integrals. To
accomplish this, we make use of the usual trick of letting
$\tilde{\omega}_{m,l}^{b,a}$ and $\tilde{\omega}_{n,l}^{c,a}$ have a
small negative imaginary part, whereby the integrals converge.
\begin{eqnarray}
\left|\psi_{c_{}}^{(3)}(\tau) \right\rangle_t &=& e^{i\theta_{k_{}}}
\sum_{k_{},l_{},m_{}}
\mathfrak{D}_{m_{},l_{}}^{b_{},a_{}}\mathfrak{D}_{n_{},m_{}}^{c_{},b_{}}
\int_{-\infty}^{\infty}\hspace{-0.42cm} d\omega_2
\int_{-\infty}^{\infty} \hspace{-0.42cm} d\omega_3\,
\nonumber\\
&& \times \frac{\mathcal{E}_2(\omega_2) \mathcal{E}_3(\omega_3)\,
e^{-i(\omega_2+\omega_3 +
\tilde{\omega}_{l_{}}^{a_{}})t}}{\left(\omega_2 +
\tilde{\omega}_{l_{},m_{}}^{a_{},b_{}} \right) \left(
\tilde{\omega}_{n_{},l_{}}^{c_{},a_{}} -\omega_2 - \omega_3 \right)}
\nonumber\\
&&\times e^{i \tilde{\omega}_{l_{},k_{}}^{a_{},0} \tau}
{}_{k}\beta_{l_{}} \left|\phi_{a_{},l_{}} \right\rangle. \nonumber
\end{eqnarray}
One may note in passing that if the dominating effect of pulse~$2$
is to stimulate emission as shown in Fig.~\ref{fig:Eoversigt}, the
negative frequency components of $\mathcal{E}_2(\omega_2)$ will give
the dominating contribution to the $\omega_2$-integral.

The electric field of the $FW$ pulse is proportional to the second
time derivative of the dipole moment
\begin{eqnarray}
E_{FW}(t,\tau) &=& \sum_k {}_{k}{E}_{FW}(t,\tau) \nonumber\\
&\propto& \sum_k \frac{d^2}{dt^2}\left[{}_{k}d(t,\tau)\right]
\nonumber
\end{eqnarray}
This yields for the contribution caused by the $k$'th initially
populated level:
\begin{eqnarray}\label{eq:d30aft}
\frac{d^2}{dt^2}\left[{}_{k}d(t,\tau)\right] &=& \frac{d^2}{dt^2}\,
\left\langle {}_{k}\psi_{c_{}}^{(3)}(\tau) \big| \hat{d} \big|
{}_{k}\psi_0^{(0)}
\right\rangle_{t\hspace{-3.05cm}t}\hspace{3.05cm}.
\end{eqnarray}
Notice, that we only have to use the same $k$'th vibrational state
from the initial thermal state, since inner products with all other
terms vanish (see section~\ref{sec:stme}). Using this result, we can
find the frequency dependence of the $FW$ pulse's electric field:
\begin{eqnarray}\label{eq:Ecogb}
\frac{{}_{k}\mathcal{E}_{FW}(\omega,\tau)}{\omega^2} &\propto& {}_{k}\tilde{d}(\omega,\tau) \nonumber\\
&=& \frac{1}{{2\pi}}\int_{-\infty}^\infty \hspace{-0.39cm} dt \, e^{i\omega t} \,{}_{k}d(t,\tau) \nonumber\\
&=& \sum_l {}_{k}C^{*}_{l_{}}(\omega)\,e^{i
\tilde{\omega}_{k_{},l_{}}^{0_{},a} \tau} {}_{k}\beta_{l_{}}^{*},
\end{eqnarray}
where the mapping coefficients ${}_{k}C_{l_{}}(\omega)$ (and their
complex conjugate) depend upon the lifetime of the energy levels and
the detailed electric fields of the pulses~$2$ and~$3$:
\begin{eqnarray}\label{eq:Coeff}
{}_{k}C^{*}_{l_{}}(\omega) &=& \sum_{m_{},n_{}}
\frac{\mathfrak{D}_{l_{},m_{}}^{a_{},b_{}}
\mathfrak{D}_{m_{},n_{}}^{b_{},c_{}}
\mathfrak{D}_{n_{},k_{}}^{c_{},0}} {\omega +
\tilde{\omega}^{c_{},0}_{n_{},k_{}}}\nonumber\\
&&\times \int_{-\infty}^{\infty} \hspace{-0.39cm} d\omega_2\,
\frac{\mathcal{E}^*_2\hspace{-0.03cm}\left(\omega_2\right)\,
\mathcal{E}^{*}_3\hspace{-0.07cm}\left(\tilde{\omega}_{k_{},l_{}}^{0_{},a}
-\omega_2 - \omega\right)}{\omega_2 +
\tilde{\omega}_{l_{},m_{}}^{a_{},b_{}}}.
\end{eqnarray}
This expression clearly shows the transitions $a \rightarrow b
\rightarrow c \rightarrow 0$ through the $\mathfrak{D}$-factors and
the resonance frequencies in the $\omega$-dependent denominators.
The apparent divergencies are not real, but a usual artifact of
applying perturbation theory.

The most important point for our purposes are the linearity of
Eq.~(\ref{eq:Ecogb}) in our sought coefficients
${}_{k}\beta_{l}^{*}$, and the exponential $\tau$-dependence.

The measurements at our disposal are spectral intensities for each
delay $\tau$, i.e. a two-dimensional spectrogram. Even though we
measure intensities and not electric fields, the signal is found
from squaring the electric field $S(\omega,\tau) \propto
|\mathcal{E}_{FW}(\omega,\tau)|^2$ \footnote{This surprising result
is a many-particle effect - a direct consequence of the vast number
of uncorrelated molecules emitting the $FW$-pulse
\cite{polbeatgofork}, \cite{Larsdrop1}.}. Therefore, dividing the
signal by $\omega^4$, we can find the following quantities:
\begin{eqnarray}\label{eq:bcbc}
\frac{S(\omega,\tau)}{\omega^4} &=& N \left|\tilde{d}(\omega,\tau)\right|^2\nonumber\\
&=& N \sum_{\substack{l,l'\\k,k'}} {}_{k}C_{l_{}}(\omega)\,
{}_{k'}C^{*}_{l'_{}}(\omega)\, {}_{k}\beta_{l_{}}\,
{}_{k'}\beta_{l'_{}}^{*} \nonumber\\
&& \hspace{0.9cm} \times e^{i
\left(\tilde{\omega}_{l,l'}^{a_{},a_{}}-\tilde{\omega}_{k_{},k_{}'}^{0,0}\right)
\tau},
\end{eqnarray}
where the overall positive factor $N$ is introduced for two reasons:
First, we do not know the absolute populations created by the
pulses, and second, we wish to avoid the difficulties connected with
measuring absolute signal strengths. To extract the
${}_{k}\beta_l$'s, a first step is to isolate single terms
${}_{k}C_{l_{}}(\omega)\, {}_{k}\beta_{l_{}}\,
{}_{k'}C^{*}_{l'_{}}(\omega)\, {}_{k'}\beta_{l'_{}}^{*}$ from the
sum in Eq.~(\ref{eq:bcbc}). We can use Fourier transformation with
respect to $\tau$ to extract a single or a few terms of this type.
The number of terms found from such a procedure equals the number of
recurrences in the transition frequencies at the chosen frequency
$\Omega$:
\begin{eqnarray}\label{eq:ftafbcbc}
F(\Omega,\omega) &=&
\frac{1}{2\pi}\int_{-\infty}^{\infty}\hspace{-0.37cm} d\tau\,
\frac{S(\omega,\tau)}{\omega^4}
 \, e^{-i \Omega \tau} \nonumber\\
&=& N \frac{1}{2\pi}\int_{-\infty}^{\infty}\hspace{-0.37cm} d\tau\,
\left|\tilde{d}(\omega,\tau)\right|^2
 \, e^{-i \Omega \tau} \nonumber\\
&=& N \sum{}' \,{}_{k}C_{l_{}}(\omega)\,
{}_{k}\beta_{l_{}}\,{}_{k'}C_{l_{}'}^{*}(\omega)\,{}_{k'}\beta^*_{l_{}'},
\end{eqnarray}
with the primed sum running over indices $(k,k',l,l')$ so that
$\Omega =
\tilde{\omega}_{l,l'}^{a_{},a_{}}-\tilde{\omega}_{k_{},k_{}'}^{0,0}$.
Even though the $\tau$-integral in Eq.~(\ref{eq:ftafbcbc}) runs over
all values of $\tau$, one cannot achieve this in real experiments.
Both the inevitable truncation of the $\tau$-interval and decay
processes will cause a broadening of the peaks in $F(\Omega,\omega)$
as a function of $\Omega$, in turn leading to an increased number of
frequency recurrences. Although a concern in principle, it will be
seen that it is possible in typical experiments to use (a discrete
version of) Eq.~(\ref{eq:ftafbcbc}) to find sums with just a single
or a few terms. We will aim to extract single terms from these sums.
Obviously, choosing $\Omega = 0$ gives many terms, among them the
ones with $l = l'$, making it difficult to extract any single one.
Fortunately, it will turn out that we will not need these, so we
shall henceforth look for terms with $l \neq l'$.

To select out a single or a few terms from the function
$F(\Omega,\omega)$ in Eq.~(\ref{eq:ftafbcbc}) we must make
appropriate choices of the variables $\Omega$ and $\omega$. In the
following subsection~\ref{sec:caseI}, we shall present a method to
do this for the $k$ that can be selected by proper choice of
$\omega$: case~(I). This is usually the case for the lowest
populated (ground) state $k = 0$ (see e.g. \cite{Runeogalbert}) and
for any $k$ that are energetically separated from their neighboring
vibrational levels by more than the combined width of the laser
pulses. Further below, in subsection~\ref{sec:Interfmelk} we will
present how to deal with the remaining $k$ through making a weak
assumption on the spectrum - case~(II). Finally, in
subsection~\ref{sec:caseIII} we will discuss how to do the
reconstruction if the two above methods fail.

\subsection{Emission due to single $k$'s: Case~(I) \label{sec:caseI}}
\begin{figure}[htbp]
\includegraphics[width=0.48\textwidth]{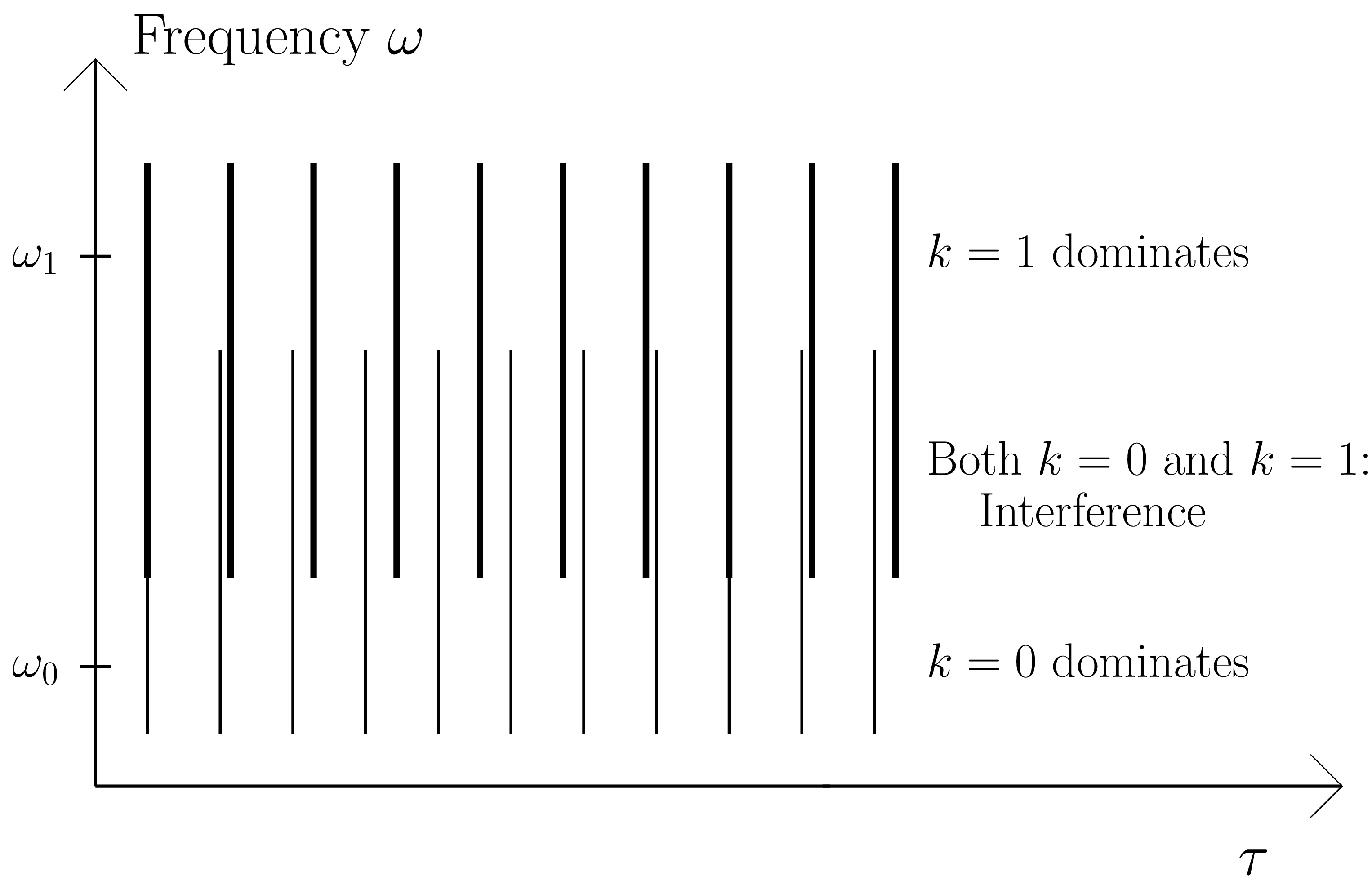}
\caption{Schematic illustration of a spectrogram Eq.~(\ref{eq:bcbc})
in case (I): There are regions of $\omega$ where only a single $k$
contributes. The vertical lines signify maxima in the signal
$S(\omega,\tau)/\omega^4$, with contributions from different initial
levels explicitly shown. To extract the numbers ${}_k\beta_l$ from
the sum Eq.~(\ref{eq:ftafbcbc}), one can choose $\omega$ to select
$k$. In the example shown in this figure, one can extract
information about ${}_{0}\beta_l$ from using the data with $\omega =
\omega_0$. Similarly, fixing the frequency at $\omega = \omega_1$
allows one to obtain the coefficients ${}_{1}\beta_l$.
\label{fig:spekskema}}
\end{figure}

We turn our focus to the group of $k$ states that belong to
case~(I). Choosing the frequency $\Omega =
\tilde{\omega}_{l,l'}^{a,a}$, we see that the sum
Eq.~(\ref{eq:ftafbcbc}) contains terms with all $k = k'$. Even
though this is a complication, we can circumvent it in this case by
using the freedom in choosing $\omega$ to select $k$ by making all
${}_{k}C_l(\omega)$ vanish, except for a single $k$. This situation
is illustrated in Fig.~\ref{fig:spekskema}.

In the usual experimental situation where the signal
$S(\omega,\tau)$ is simultaneously recorded for all frequencies
$\omega$ in a wide range, we also know $F(\Omega,\omega)$ for this
range of $\omega$. To find out if only one $k$ contributes at a
fixed $\omega$, one simply checks whether
$F(\tilde{\omega}^{a,a}_{l,l'}-\tilde{\omega}^{0,0}_{k,k'},\omega) =
0$ for all $k' \neq k$.

To avoid testing too many values of $\omega$ by the above procedure,
one can readily determine the approximate $\omega$-range of
interest. From the easy measurements of the central frequency and
widths of pulses~$1$-$3$, one can locate a limited $\omega$-interval
where one expects the $FW$ signal from each $k$. In such
considerations, one can also benefit from knowing approximate
potential surfaces to get an idea of which transitions have
appreciable transition matrix elements.

Thus, by wisely selecting certain values of $\omega$, we can limit
the primed sum Eq.~(\ref{eq:ftafbcbc}) to run over only $l$ and
$l'$. If there are degeneracies in the frequencies
$\tilde{\omega}_{l,l'}^{a,a}$, it can still be possible to find the
elements indirectly. Since the details depends on the problem at
hand, we illustrate this with an example.

Say, $\tilde{\omega}_{l_1,l_{1}'}^{a,a} =
\tilde{\omega}_{l_2,l_{2}'}^{a,a}$, and we have Fourier transformed
to find the following sum, suppressing the $k$-index and
$\omega$-variable for clarity:
\begin{eqnarray}\label{eq:felled}
F(\Omega) &=& N C_{l_{1}}\beta_{l_{1}} C_{l_{1}'}^{*}
\beta^*_{l_{1}'} + N C_{l_{2}} \beta_{l_{2}}
C_{l_{2}'}^{*}\beta^*_{l_{2}'}.
\end{eqnarray}
Provided that we can find each of the following coefficients and
that none of them are zero, we can form the quotient:
\begin{eqnarray}
\frac{N C_{l_1}\beta_{l_1}C_{l_3}^*\beta_{l_3}^* \times N
C_{l_3'}\beta_{l_3'}C_{l_{1}'}^*\beta_{l_{1}'}^*}{N
C_{l_3'}\beta_{l_3'}C_{l_3}^*\beta_{l_3}^*} &=& N
C_{l_{1}}\beta_{l_{1}}C_{l_{1}'}^{*} \beta^*_{l_{1}'},\nonumber
\end{eqnarray}
allowing us to extract each of the terms in Eq.~(\ref{eq:felled}).
Thus, we can use intermediate levels to obtain each term in sums,
otherwise hidden because of the degeneracy in the transition
energies.

In principle, we could similarly find all such products $N
C_{l_{}}\beta_{l_{}}C_{l_{}'}^{*}\beta^*_{l_{}'}$, including $l =
l'$. Though having all these products permits a nice statistical
treatment through singular value decomposition of the outer product
matrix, it can be difficult to measure $N
C_{l_{}}\beta_{l_{}}C_{l_{}'}^{*}\beta^*_{l_{}'}$ with the integers
$l$ and $l'$ differing by more than a few. This is because terms
with $l$ and $l'$ significantly different correspond to quantum
beats between distant levels in the unknown excited state.
Fortunately, these are not necessary because of our freedom of
choosing normalization and phase of each $\left|{}_{k}\psi_{a}^{(1)}
\right\rangle$, as will be clear below.

We have now reached the limit of what we can extract from the
$(\omega,\tau)$-spectrograms. To find the unknown quantum state
through the $\beta_{l_{}}$'s, we need to know the
$C_{l_{}}(\omega)$'s. These are in turn both difficult to calculate,
one reason being the regularization of the apparent divergencies in
Eq.~(\ref{eq:Coeff}), and more importantly require detailed
knowledge of the electric fields of the pulses~$2$ and $3$, of the
dipole matrix elements $\mathfrak{D_{\nu,\nu'}^{\alpha,\alpha'}}$
and of the transition frequencies
$\omega_{\nu,\nu'}^{\alpha,\alpha'}$. One may rightly argue that
with the knowledge of the dipole transition moments, it would have
been easier to measure the field of the single pulse~$1$ and simply
calculate the unknown quantum state by propagation using the
dynamical laws. Therefore, we shall take a completely different
approach which circumvents both regularizations, detailed knowledge
of electric fields, and knowledge of transition matrix elements and
-energies between excited states.

\subsection{Calibration pulse\label{sec:calpuls}}
In this subsection, we introduce one or more calibration pulses
aimed at creating excited quantum states in the electronic level $a$
with known calibration expansion coefficients
${}_{k}\beta^{\textnormal{cal}}_l$. This is a different approach
than heterodyne detection or other reference pulse techniques, since
the calibration experiment is a separate experiment from the one
where the reconstruction data are recorded. Hence, by performing
exactly the same experiments as above, except with the calibration
pulse substituted for pulse~$1$, we will find products of the form
$N_{C}\,{}_{k}C_{l_{}}(\omega)\, {}_{k'}C_{l_{}'}^{*}(\omega)$.
These will in turn be used in the reconstruction experiment with
pulse~$1$ to find the coefficients ${}_{k}\beta_l$. Let us consider
this in detail.

Contrary to pulse~$1$, we will require that the calibration pulse(s)
be perturbative, because this gives simple expressions for the state
created in the calibration. We find this state similarly to
Eq.~(\ref{eq:psitopert}), letting the upper limit in the integration
tend to infinity since the electric field
$E_{\textnormal{cal}}(t'+\tau)$ vanishes for times long before the
pulses~$2$ and $3$ are turned on:
\begin{eqnarray}\label{eq:calpsi}
\left|{}_{k}\psi_a^{\textnormal{cal}}(\tau)\right\rangle_{t_2} &=&
-\frac{iq}{\hbar} e^{i\theta_k}\int_{-\infty}^{\infty}
\hspace{-0.37cm}dt'\, e^{-i \hat{H}_{a}(t_{2}-t')/\hbar}
E_{\textnormal{cal}}(t'-\tau)\nonumber\\
&&\qquad \qquad \times \,\hat{d}\, e^{-i\hat{H}_{0}(t'-0)/\hbar}
\left|{}_{k}\psi^{(0)}_{0}\right\rangle_{0} \nonumber \\
&=& -i\,e^{i\theta_k} \sum_l \mathfrak{D}_{l,k}^{a,0}\,
\mathcal{E}\hspace{-0.07cm}\left(\tilde{\omega}_{l,k}^{a,0}\right)\,
\left|\phi_{a,l} \right\rangle\nonumber\\
&& \qquad \qquad \times \, e^{i
\tilde{\omega}_{l,k}^{a,0}\tau}e^{-i\tilde{\omega}_{l}^{a}t_2}.
\end{eqnarray}
Comparing Eq.~(\ref{eq:calpsi}) to Eq.~(\ref{eq:ukendttilst}), we
see that the expansion coefficients of the calibration state are
\begin{eqnarray}
{}_{k}\beta^{\textnormal{cal}}_l = - i \mathfrak{D}_{l,k}^{a,0}\,
\mathcal{E}\hspace{-0.07cm}\left(\tilde{\omega}_{l,k}^{a,0}\right).
\end{eqnarray}
Thus, by knowing the dipole transition elements
$\mathfrak{D}_{l,k}^{a,0}$ and the frequency-components of the
electric field of a calibration pulse, we can calculate the
expansion coefficients of the vibrational calibration state created
in the electronic state $a$. For convenience, one can merely use a
calibration pulse that has a constant spectrum over the relevant
frequencies $\tilde{\omega}_{l,k}^{a,0}$, giving calibration
expansion coefficients of the levels according to the values of the
dipole transition matrix elements. It is therefore not necessary to
have detailed information about the calibration pulses, but we must
know the transition dipole matrix elements between the electronic
states $0$ and $a$. Fortunately, these are typically readily
measured, e.g. by fluorescence spectroscopy, or calculated. In
contrast, it is more difficult to find similar elements between two
excited states - elements we fortunately do not need to know.

In the remaining part of this subsection we will be considering a
single $k$, and we shall suppress this index for clarity.

For the calibration to be useful, the calibration data recorded must
enable us to determine the $\beta_l$'s up to a common non-zero
complex factor. We will accomplish this by determining products
$\beta_{l} \beta^{*}_{l'}$ up to an overall complex factor, common
for all products. The measurements with pulse~$1$ gives quantities
$N_{1} C_{l}(\omega) \beta_{l}\,C_{l'}^{*}(\omega) \beta^{*}_{l'}$,
precisely allowing us to determine $N_{1} \beta_{l} \beta_{l'}^{*}$
if we know $C_{l}(\omega) C_{l'}^{*}(\omega)$. These mapping
coefficient products can in turn be found, up to a common factor
$N_{C}$, from the calibration data $N_{C} C_{l}(\omega)
\beta^{\textnormal{cal}}_{l}\,C_{l'}^{*}(\omega)
\beta^{\textnormal{cal},*}_{l'}$, where the
$\beta_l^{\textnormal{cal}}$'s are known. By division, we can hence
find products of the form $N' \beta_{l} \beta_{l'}^{*}$. To
determine the $\beta_{l}$'s individually, we simply set one of these
equal to one, use the products to determine the rest, and finally
normalize. Therefore, there is no benefit in knowing one of these
products where neither $l$ or $l'$ enter in another product.
Consequently, we must know the range of $l$'s that are populated by
pulse~$1$ (from its spectral center and width) and be able to find
products $N' \beta_{l} \beta^{*}_{l'}$ where all populated $l$'s
enter, and no product has both an $l$ and an $l'$ that do not enter
in another. Furthermore, one is free to use data from more than one
value of $\omega$, as long as each data set can be related to data
at all other values of $\omega$. This is sufficiently fulfilled if
one can determine a product $N_{C} C_{l}(\omega) C_{l'}^{*}(\omega)$
at one value of $\omega$ that contains an $l$ also contained at in a
product at another value of $\omega$.

In case it is not convenient to populate all these levels with a
single calibration pulse, more can be used in subsequent
calibrations. Since we can allow for only a single undetermined
overall complex factor, it is necessary to relate the data from one
calibration pulse to the others. For instance, it is sufficient to
determine a single $N_{C1} C_{l}(\omega) C^{*}_{l'}(\omega)$ from
one calibration $C1$ which can also be determined in another
calibration $C2$ to find the ratio $N_{C1}/N_{C2}$.

All this having been said, we illustrate the procedure with a brief
example. Let us say that we are interested in reconstructing the
vibrational state originating in a certain $k$ (which we suppress
for clarity), and we desire the expansion coefficients $\beta_l$
with $l \in \{1,2,3,4,5\}$. We employ two calibration pulses~$C1$
and $C2$. $C1$ has a constant spectrum in a range sufficient to
populate levels $1$-$4$ and $C2$ similarly populates levels $3$-$5$.
Measuring at the two frequencies $\omega_a$ and $\omega_b$ where
other $k' \neq k$ give only negligible contributions, it is found
that there is good signal strength for the following terms:
\begin{eqnarray*}
\left. \begin{array}{l}
N_{C1}\, C_1(\omega_a)C_{2}^{*}(\omega_a)\\
N_{C1}\, C_2(\omega_a)C_{4}^{*}(\omega_a)\\
N_{C1}\, C_3(\omega_b)C_{4}^{*}(\omega_b)
\end{array} \right\}\textnormal{pulse }C1\\
\left. \begin{array}{l}
N_{C2}\, C_3(\omega_b)C_{4}^{*}(\omega_b)\\
N_{C2}\, C_3(\omega_b)C_{5}^{*}(\omega_b)\\
\end{array} \right\}\textnormal{pulse }C2\\
\end{eqnarray*}
where we have already divided out the known calibration expansion
coefficients $\beta^{\textnormal{cal}}_{l}
\beta^{\textnormal{cal},*}_{l'}$, up to a factor absorbed in the
$N_{C}$'s. The element $C_3(\omega_b)C^{*}_4(\omega_b)$ recurring in
both pulses serves the special purpose of relating the two pulses by
showing us the ratio of the numbers $N_{C1}$ and $N_{C2}$.
Hereafter, the actual experiment is performed with pulse~$1$. In the
table below, we show how one obtains the products $N' \beta_l
\beta_{l'}^{*}$ from measurements and calibration data.
\\
\\
\begin{tabular}[c]{l|c|r}
Measurement: & Used calibration: & We find:\\
\hline \hline
$N_1\, C_1(\omega_a)\beta_1 C_2^{*}(\omega_a)\beta_2^{*}$&$N_{C1}\, C_1(\omega_a)C_2^{*}(\omega_a)$&$N'\beta_1 \beta_2^{*}$\\
\hline
$N_1\, C_2(\omega_a)\beta_2 C_4^{*}(\omega_a)\beta_4^{*}$&$N_{C1}\, C_2(\omega_a)C_4^{*}(\omega_a)$&$N'\beta_2 \beta_4^{*}$\\
\hline
$N_1\, C_3(\omega_b)\beta_3 C_4^{*}(\omega_b)\beta_4^{*}$&$N_{C1}\, C_3(\omega_b)C_4^{*}(\omega_b)$&$N'\beta_3 \beta_4^{*}$\\
\hline
$N_1\, C_3(\omega_b)\beta_3 C_4^{*}(\omega_b)\beta_4^{*}$&$N_{C2}\, C_3(\omega_b)C_4^{*}(\omega_b)$&$N'\beta_3 \beta_4^{*}$\\
\hline
$N_1\, C_3(\omega_b)\beta_3 C_5^{*}(\omega_b)\beta_5^{*}$&$N_{C2}\, C_3(\omega_b)C_5^{*}(\omega_b)$&$N'\beta_3 \beta_5^{*}$\\
\hline
\end{tabular}
\\
\\

Since we have the freedom of choosing the normalization constant
$N'$, we can arbitrarily set $\beta_1 = 1$, from which we can use
the following lines in the table to find (in order) $N' \beta_2$,
$N' \beta_4$, $N' \beta_3$, and from the last line $N' \beta_5$.
Finally, the state can be normalized by appropriately choosing $N'$.
We have thus found the set of complex numbers $\{{}_{k}\beta_l\}$,
both their magnitude and phase, and we have thereby performed the
reconstruction of the state formed by pulse~$1$.

Having been concerned with how to use small sets of measurements, it
should be said that in a typical experimental situation, one will
have much more data than is minimally required. This abundance of
data can be used to further improve the statistics of the
reconstructed values.

Finally, it may be of note that we can easily find the individual
mapping coefficients $C_{l}(\omega)$ from the calibration data, up
to a common factor (rather than just products of two coefficients).
This is done in the exact same way as finding the
$\{{}_{k}\beta_l\}$, i.e. by setting one of the $C_{l}(\omega) = 1$,
and using the products to find the rest. Thereby dynamical
information about the molecule can be obtained.

\subsection{Interference between different $k$: Case (II)\label{sec:Interfmelk}}
\begin{figure}[htbp]
\includegraphics[width=0.48\textwidth]{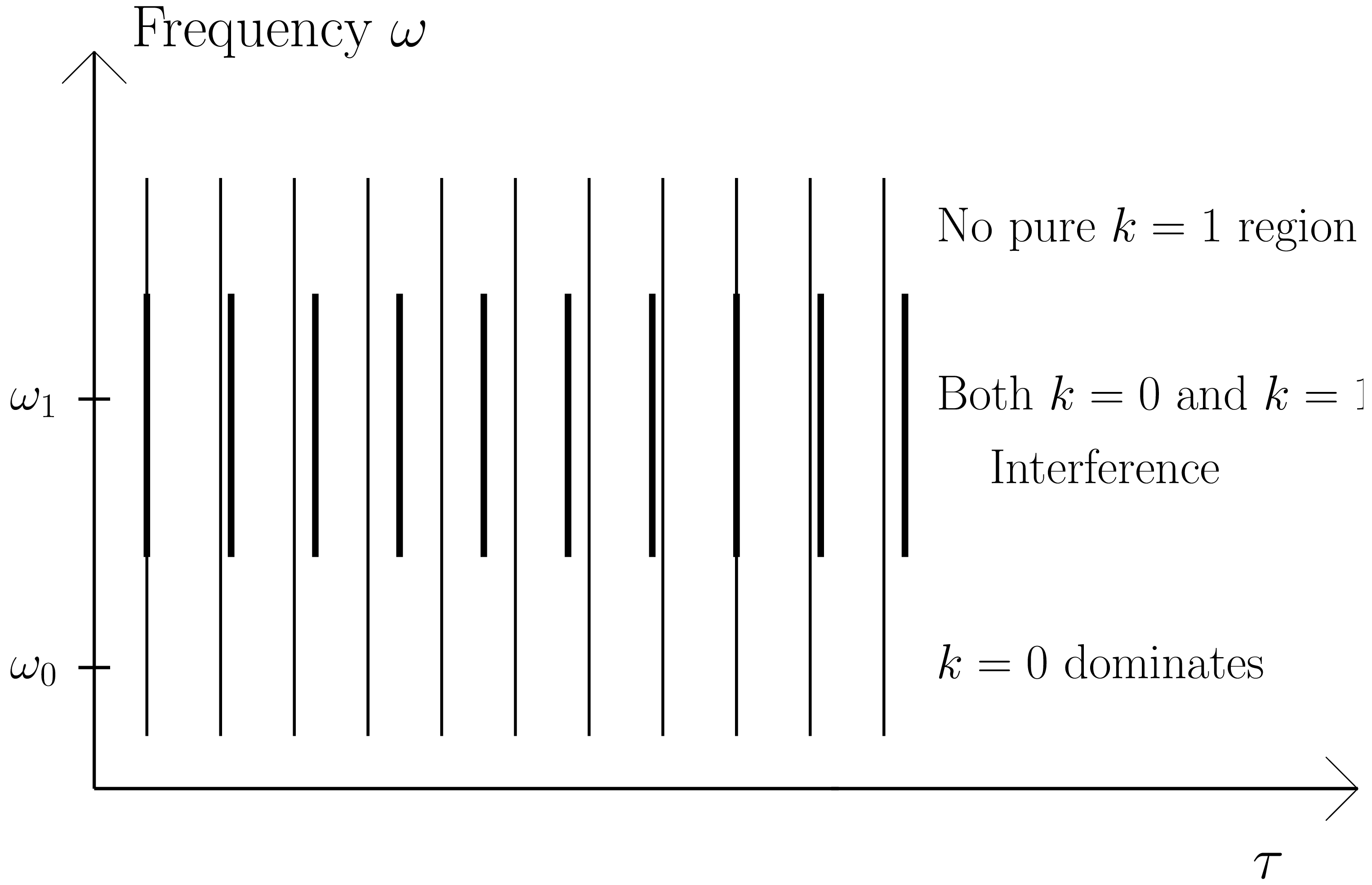}
\caption{Schematic illustration of a spectrogram Eq.~(\ref{eq:bcbc})
in case~(II). The vertical lines signify maxima in the signal
$S(\omega,\tau)/\omega^4$, with contributions from different initial
levels explicitly shown. In contrast to Fig.~\ref{fig:spekskema}, we
cannot find values of $\omega$ for which only $k = 1$ contributes.
Still, one can first find the ${}_{0}\beta_l$'s through the
procedure described in section~\ref{sec:calpuls}. These coefficients
can afterwards be used to extract the ${}_{1}\beta_l$ by using the
frequencies in $F(\Omega,\omega)$ that arise from optical
interferences between the $k = 0$ and $k' = 1$ signals. If there are
more overlapping $k'$ signals than the two shown, the
${}_{k'}\beta_l$ from these can be extracted by similar means.
\label{fig:spekskema2}}
\end{figure}

We turn our attention to case~(II) where not all $k$ can be isolated
from Eq.~(\ref{eq:ftafbcbc}) by proper choice of $\omega$. This
situation, encountered in \cite{Runeogalbert}, is illustrated in
Fig.~\ref{fig:spekskema2} for two initially populated $k$-levels.
Here the wave-packet originating in $k = 1$ cannot be found by the
above procedure.

However, if just a single $k$ can be isolated by choosing $\omega =
\omega_0$, we can by the above method find the ${}_{0}\beta_l$'s for
this $k = 0$ (usually the ground state) up to a common phase factor.
In this subsection we describe how the ${}_{k \neq 0}\beta_l$ can be
found if the spectrum fulfills certain conditions. For transparency,
we initially focus on the situation in Fig.~\ref{fig:spekskema2}
where we seek the coefficients ${}_{1}\beta_l$. For our method to
work, we demand that some of the frequencies $\Omega^{0,1}_{l,l'} =
\tilde{\omega}_{l,l'}^{a_{},a_{}}-\tilde{\omega}_{k = 0,k' =
1}^{0,0}$, arising from optical interferences between $k = 0$ and $k
= 1$, must be spectrally resolved from the pure $k = 0$ and $k = 1$
signals having $\Omega^{k,k}_{l,l'} =
\tilde{\omega}_{l,l'}^{a_{},a_{}}$. Having this requirement
fulfilled, we can extract the quantities $
F(\Omega^{0,1}_{l,l'},\omega_1) =
{}_{0}C_{l_{}}(\omega_1)\,{}_{0}\beta_{l_{}}\,{}_{1}C_{l_{}'}^{*}(\omega_1)\,{}_{1}\beta^*_{l_{}'}$.
In the calibration experiment, we can hence determine the mapping
coefficient product
${}_{0}C_{l_{}}(\omega_1)\,{}_{1}C_{l_{}'}^{*}(\omega_1)$ up to a
common multiplicative factor and, by using the ${}_{0}\beta_l$ found
in the experiment (using $\omega = \omega_0$), we can find
${}_{1}\beta_l$ (using $\omega = \omega_1$).

Similarly to case~(I), we do not require spectral resolution of all
frequencies $\Omega^{0,1}_{l,l'}$ and $\Omega^{k,k}_{l,l'}$. Indeed,
to determine all ${}_{1}\beta_l$, we just need one resolvable
$\Omega^{0,1}_{l,l'}$ for each $l$ at some $\omega_1$, and that the
corresponding ${}_{0}\beta_l'$ is non-zero.

If more than two values of $k$ overlap, these can be found through
similar means; either by their interference with $k = 0$ or
recursively, i.e. $k+1$ is found from $k$. In conclusion, we can
completely reconstruct the state in case~(II).

\subsection{Case~(III) \label{sec:caseIII}}
The treatments of cases~(I) and (II) above were based on the
assumption that at least a single $k$ can be isolated from
$F(\Omega,\omega)$ by appropriately choosing $\omega$. If such an
$\omega$ cannot be found, the situation is radically different.
Still maintaining that we wish to avoid making assumptions about the
mapping coefficients ${}_{k}C_l(\omega)$, one solution is to perform
several experimental runs at different temperatures of the initial
state. There are two main effects that change the spectrograms as
the temperature is varied: Temperature dependent decoherence
processes and thermal population of the initial state. The most
important decoherence process will usually be collisional
decoherence (pressure broadening), with a smaller contribution from
Doppler broadening. In any case, these two effects broaden the
spectral lines, scaling as $\sqrt{T}$ for fixed particle density.
For moderate temperature ranges, this is only a small effect; indeed
we can ignore it altogether as long as the required spectral
resolution is not washed out. In contrast, the initial state
populations scale exponentially with $1/T$, greatly changing the
spectrogram. It is exactly this change we can use to our advantage.
Thus, the terms in Eq.~(\ref{eq:ftafbcbc}) having $k = k'$ will
scale differently with temperature for each $k$, and differently
from terms with $k \neq k'$. Demanding here that pulse $1$ be
perturbative, these variations with temperature simply follows the
Boltzmann distribution corresponding to the energy of the $k$'th
level in the initial thermal state. By forming linear combinations
of $F(\Omega,\omega)$ recorded at different temperatures with
weights calculable from the corresponding Boltzmann distributions,
one can find each of the individual terms with $k = k'$ and with $k
\neq k'$. For example, if $k_1$ and $k_2$ both enters in
Eq.~(\ref{eq:ftafbcbc}), one can find all the terms individually;
i.e. terms containing $(k_1,k_1)$, $(k_2,k_2)$ and $(k_1,k_2)$.
Consequently, it is possible to modify the above reconstruction
method so it can be used, even if all $k$ cannot be selected
individually by choosing special values of $\omega$.

\section{Discussion\label{sec:disk}}
Having presented above a specific implementation of the
reconstruction method, we will now discuss how to do away with the
clarifying, but non-essential, assumptions we made in
section~\ref{sec:experim}. We discuss first the effects of rotation
in subsection \ref{sec:rotat}, whereafter we will move on to discuss
the possible generalizations of the method in subsection
\ref{sec:generaliz}.

\subsection{Rotations \label{sec:rotat}}
In the above, we completely neglected rotations, the qualitative
correctness of which we will now justify. Though of general
validity, we will for concreteness discuss diatomic molecules with
rotational energy $E_{rot} = J(J+1)$, where $J$ is the angular
momentum quantum number. If rotational states up to a certain
${J}_{max}$ are appreciably populated, the number of rotational
states involved is circa $({J}_{max}+1)^2$. In an experiment where
the temperature is high enough to populate several vibrational $k$
states, the number of entering rotational states will be large - of
the order $N_{rot} \approx 10^4$ for the lowest vibrational state in
the experiment \cite{Runeogalbert}. All these ro-vibrational levels
are initially incoherent, and we can label each of them after their
initial quantum number $J$ and its projection $M$; The initial state
is therefore incoherent in all of the indices $k, J$ and $M$. Thus,
by analogy with Eq.~(\ref{eq:d30aft}), the dipole moment leading to
coherent $FW$ emission only contains contributions of the form
${}_{k,J,M}d(t,\tau) =\, \langle{}_{k,J,M}
\psi^{(3)}_{c}(\tau)|\hat{d}|{}_{k,J,M}\psi^{(0)}_0 \rangle_{t
\hspace{-4.13cm}t\hspace{4.11cm}}$. For linearly polarized pulses,
each of the one-photon transitions shown in Fig.~\ref{fig:Eoversigt}
changes $J$ by $0$ or $\pm 1$ while $M$ is conserved. Whether
$\Delta J = 0$ is allowed depends on the symmetry of the electronic
states $0-c$. In the diatomic case, $\Delta J = 0$ requires one of
the involved states in a one-photon transition to have non-zero
electronic angular momentum. As a consequence, all the transition
energies between vibrational levels $l$ and $l'$ entering in
Eq.~(\ref{eq:bcbc}) (for clarity, with $k = k'$) will be split into
many $\tilde{\omega}^{a,a}_{(l,J_1,M),(l',J_2,M)}$ with $J_1$ and
$J_2$ differing by at most two. As known from field free alignment,
some of these terms will interfere destructively after a time
$\mathcal{T}_{di} = \pi\hbar/2B\bar{J}$ and have a full revival
after $\mathcal{T}_{rev} = \pi\hbar/B$ where $\bar J$ is the mean
value of $J$ \cite{Vrakking1}. To elucidate which terms interfere
destructively, one can write down the frequency dependent dipole
moment for fixed $k$ and $l$ quantum numbers in the schematic form
(i.e. neglecting the difference in populations and mapping
coefficients) $\bm{d} \sim
\sum_{J,M}[\bm{d}_{JM,0}+\bm{d}_{JM,+1}+\bm{d}_{JM,-1}]$, where
$\bm{d}_{JM,\Delta J}$ is the dipole moment contribution originating
in the state with angular momentum quantum number $J$ and having
this changed by $\Delta J$ in the first transition. Finding the
signal $S(\omega) \propto |\bm{d}|^2/\omega^4$, we find
$N_{rot}(N_{rot}+2) \approx N_{rot}^2$ terms of the form
$|\bm{d}_{J,\Delta J}|^2$ with identical $\tau$-dependence; the
majority ($N_{rot}^2$) coming from $|\bm{d}_{J,\Delta J=0}|^2$.
These terms will all stay in phase and give rise to a signal at all
delay times $\tau$, scaling as $N_{rot}^2$. The remaining $8
N_{rot}^2-2N_{rot} \approx 8 N_{rot}^2$ terms will have different
$\tau$-dependent phases, and will be out of phase a time
$\mathcal{T}_{di}$ after formation by pulse~$1$. This makes their
contribution to the signal scale as $N_{rot}$, except at the
fractional and full revivals. Summarizing, for delays greater than
$\mathcal{T}_{di}$, but smaller than major fractional revivals, the
terms with $\Delta J = 0$ will completely dominate the signal due to
the scaling $N_{rot}^2 \gg N_{rot}$. Because all these terms have
the same $\tau$-dependence, one can completely ignore the rotational
degrees of freedom. A similar argument can be applied to pulse~$2$
and $3$ showing that to avoid diminishing the signal further due to
rotational interference (i.e. loose about $8/9$ of the signal), it
is beneficial to send in pulse~$2$ and $3$ so close in time that
approximately no rotational evolution takes place during their
separation time. Having accounted for the scalings of the different
contributions to the signal, we must also account for the possible
delay times in an experiment. Due to collisional decoherence, it is
typically only possible to measure the spectrogram for values of
$\tau$ much smaller than the rotational revival time. Conversely,
the vibrational periods are short enough to go through many cycles
before the signal is washed away by decoherence. In conclusion, it
is qualitatively correct to ignore the rotational degrees of freedom
as long as (i) $N_{rot} \gg 1$, (ii) $\Delta J = 0$ transitions are
allowed, (iii) we limit the $\tau$-interval for the Fourier
transform in Eq.~(\ref{eq:ftafbcbc}) so that pulse~$1$ ends more
than $\mathcal{T}_{di}$ before pulse~$2$ and $3$, and (iv) we avoid
values of $\tau$ so that pulse~$2$ arrives at a major fractional
rotational revival.

The rotations also introduces ro-vibrational coupling, which
effectively contributes an additional line-broadening. Though
usually negligible, this effect scales with the absolute temperature
as $T$ - more rapidly than pressure and doppler broadening. What
limits the usable temperature is that the required spectral
resolution may not be washed out by these line broadenings.

Lastly, it was suggested in section~\ref{sec:caseIII} to perform
experiments at different temperatures, altering both the vibrational
and rotational populations. Being unable to resolve the individual
rotational states within a vibrational level, we have taken all
these rotational sub-levels to be included in their common
vibrational level. In this way, all mapping coefficients contain
sums over rotational levels and it is necessary to consider how
these mapping coefficients change with temperature due to the
changed rotational content. There could be two causes of such a
change. Firstly, the transition frequencies
$\tilde{\omega}_{\nu,J,\nu',J'}^{\alpha,\alpha'}$ in
Eq.~(\ref{eq:Coeff}) depend on $J$. Since $\Delta J = 0, \pm 1$ in
each transition, these differences in transition frequencies are
rather small and may be negligible. If not, one can merely separate
pulses $2$ and $3$ by more than $\mathcal{T}_{di}$, ensuring that
only $\Delta J = 0$ is measured whereby the transition frequencies
are independent of $J$. Secondly, it must be clarified how the
mapping coefficients and unknown coefficients depend on the
temperature through their rotational states. Separating the
ro-vibronic states in a product of a rotational and vibronic part,
this is straightforward, but cumbersome. An easy situation arises,
however, for linear molecules where many rotational states are
populated. Here, the electronic angular momentum in all states $0-c$
is of the order $1$ and the main contribution to the signal comes
from $J \gg 1$. Then, the transitions depend only slowly on $J$, and
the change in difference in rotational distributions with
temperature may be ignored altogether.

\subsection{Generalizations of the method \label{sec:generaliz}}
Having dealt with rotations, we turn to relaxing the other
assumptions made in section~\ref{sec:experim} and the possible
generalizations of the method. All of these generalizations are
centered around the use of calibration and around the linearity in
expansion coefficients arising because of the perturbative character
of the pulses~$2$ and $3$, see e.g. Eq.~(\ref{eq:Ecogb}).

First, is easy to see that this structure will be the same
regardless of whether we are restricted to dipole transitions or
also include higher multi-pole transitions. While this may not be of
much relevance in four-wave mixing experiments, it shows the general
principle: While the mapping coefficients ${}_{k}C_l(\omega)$ become
more complicated, these are still fully accounted for because of the
use of calibration pulses.

Second, the implicit assumption made in Fig.~\ref{fig:Eoversigt} and
in section~\ref{sec:teori} that the photon from pulse~$2$ enters the
interaction before the photon from pulse~$3$ can also be relaxed.
Similarly to the incorporation of multi-pole transitions above, one
can sum over the different paths through the energy landscape traced
out by the different sequences of absorbtion of photons from the
pulses~$2$ and $3$. For instance, if there is temporal overlap of
the probe pulses~$2$ and $3$, a major contribution to the electric
field can be caused by an inner product of the form $\langle
{}_{k}\psi_{a}^{(1)}(\tau)|\hat{d}|{}_{k}\psi_{b}^{(2)}\rangle$.
Here, $|{}_{k}\psi_{b}^{(2)}\rangle$ is independent of $\tau$, and
is formed from the initial state by absorption of a photon from
pulse~$2$ followed by stimulated emission of a photon to pulse~$3$.
Like other energetic paths, this preserves the linearity in the
${}_{k}\beta_{l}$'s and the signal's dependence on $\tau$. Indeed,
one could even incorporate multi-photon transitions and still retain
the structure of Eq.~(\ref{eq:Ecogb}) and the reconstruction
procedure from section~\ref{sec:calpuls}, provided that the
truncated perturbation series is an accurate description at each
level of perturbation. If the molecular electronic levels are
conveniently spaced, one can even use another number of pulses for
the above procedure, down to a minimum of two (one for excitation
and one for probing). The special importance of the FW situation is
that it uses the smallest number of pulses in the widespread
situation where only two electronic levels $0 = b$ and $a = c$ are
used. Though our method allows for both different numbers of
incoming pulses and for multi-photon transitions, the non-co-linear
geometry still plays an important role. Its advantage lies not in
selecting certain electronic states, but rather in avoiding the
strong background from pulses~$1$-$3$. Additionally, it gives a
reasonably simple way of finding the $\omega$-interval that allows
us to select a certain $k' = k$ in Eq.~(\ref{eq:ftafbcbc}).

Third, it is unnecessary that the electronic indices $0$ and $a$-$c$
denote single electronic states. These could as well be groups of
states, where the vibrational indices $k$-$n$ simply denotes all
states in these groups. With this generalization it also becomes
apparent that we do not need to make the Born-Oppenheimer
approximation to maintain Eq.~(\ref{eq:Ecogb}). Without this
approximation, we cannot speak of vibrational and electronic states,
but we make instead the more general statement that we find the
quantum state in the group of energy levels labeled by $a$.

Fourth, the unknown state does not need to be formed by a single
laser pulse. What made it convenient to use the single laser
pulse~$1$ above, was that the BOX-configuration made it possible to
distinguish the signal from the states having interacted with
pulse~$1$ by the direction of the $FW$-pulse. Nevertheless, any
coherent process forming an unknown state in the group of levels $a$
will work, provided that there remains population in the original
group $0$ levels (i.e. the states labeled by $k$), and no population
is transferred back to these levels by the excitation process.

Fifth, the method above is easily extended to the case of correlated
initial states, rather than just thermal ones. The difference from
the above treatment is that there can now be quantum correlations
between the different $k$-levels. To treat this situation, one
simply writes down the density operator corresponding to the vector
Eq.~(\ref{eq:psitotal}), retaining the $(k,k')$-correlations.
Similarly to the treatment above, one looks for the coherent
$FW$-terms in the emitted radiation with the appropriate direction.
If one can again find $\omega$-values where the mapping coefficients
are non-zero for a single $k$ only, one can use the method above to
find the same density matrix elements that we found previously.
These are the elements arising due to the diagonal terms in the
energy basis of the initial state. The difference, compared to the
thermal case, is that there is additional information about the
state found in an $\omega$-region approximately
$\tilde{\omega}_{k,k'}^{0,0}$ from the $(k,k)$-region. This being
said, it may well be easier to find the $(k,k')$-correlations in the
initial state, instead of trying to extract them from the unknown
excited state.

Sixth and finally, there is no requirement that the different $k$
refers to the same molecule. If one wanted to examine a weak
emission from a molecule $A$, it could be possible to perform the
above experiment on a mixture of $A$ with another, more strongly
emitting molecule $B$. In the spectrogram region where the signals
from $A$ and $B$ overlap, the strong electric field from $B$ amplify
the signal from $A$ rather like the idea of mixing a weak signal
with that of a strong local oscillator in heterodyne detection. The
basic principle behind this was demonstrated by Engel \textit{et
al.} \cite{Engeletal}.

By discussing these generalizations, we hope to have given the
reader the impression that doing quantum state reconstruction by
perturbative processes with calibration is a very general approach
with wide applicability.

In conclusion, we have presented a method to reconstruct an unknown
vibrational quantum state, coherently excited from an initial
thermal state. By using one or more calibration pulses, we have
shown how to circumvent knowledge of the perturbative probing
process, and we have suggested that this idea may be of much greater
applicability in quantum state reconstruction.

\begin{acknowledgments}
A.~S.~M. wishes to thank Klaus M\o lmer, Lars Bojer Madsen and
Flemming Hegelund for useful comments. S.~G. acknowledges a
scholarship from the Deutsche Akademie der Naturforscher Leopoldina,
grant No. BMBF-LPD 9901/8-139.
\end{acknowledgments}

\end{document}